\documentclass[journal,onecolumn]{IEEEtran}
\IEEEoverridecommandlockouts

\usepackage{ifthen}
\usepackage{cite}
\usepackage{url}
\usepackage{amssymb}
\usepackage{amsmath}
\usepackage{algorithm}
\usepackage{algpseudocode}
\usepackage{color}
\usepackage{graphicx}
\usepackage{booktabs}
\usepackage{array}
\usepackage{multirow}
\usepackage{adjustbox}
\usepackage{siunitx}

\newtheorem{theorem}{Theorem}
\newtheorem{lemma}{Lemma}
\newtheorem{definition}{Definition}
\newtheorem{example}{Example}

\newtheorem{remark}{Remark}

\makeatletter
\newenvironment{breakablealgorithm}
  { %\begin{breakablealgorithm}
   \begin{center}
     \refstepcounter{algorithm}% New algorithm
     \hrule height.8pt depth0pt \kern2pt% \@fs@pre for \@fs@ruled
     \renewcommand{\caption}[2][\relax]{% Make a new \caption
       {\raggedright\textbf{\ALG@name~\thealgorithm} ##2\par}%
       \ifx\relax##1\relax % #1 is \relax
         \addcontentsline{loa}{algorithm}{\protect\numberline{\thealgorithm}##2}%
       \else % #1 is not \relax
         \addcontentsline{loa}{algorithm}{\protect\numberline{\thealgorithm}##1}%
       \fi
       \kern2pt\hrule\kern2pt
     }
  }{ %\end{breakablealgorithm}
     \kern2pt\hrule\relax% \@fs@post for \@fs@ruled
   \end{center}
  }
\makeatother

\begin{document}
\title{Proactive Coded Caching Scheme for D2D Networks}

	\author{
		Qiaoling Zhang, Changlu Lin, Minquan Cheng
		\thanks{Q. Zhang and C. Lin are with the School of Mathematics and Statistics, Key Laboratory of Analytical Mathematics and Applications (Ministry of Education), Fujian Key Laboratory of Analytical Mathematics and Applications (FJKLAMA), Center for Applied Mathematics of Fujian Province (FJNU), Fujian Normal University, Fuzhou 350117, China (e-mail:  qlzhang2017@hotmail.com, cllin@fjnu.edu.cn).}
		\thanks{M. Cheng is with the Key Lab of Education Blockchain and Intelligent Technology, Ministry of Education, and also with the Guangxi Key Lab of Multi-source Information Mining $\&$ Security, Guangxi Normal University, 541004 Guilin, China (e-mail: chengqinshi@hotmail.com).}
	}
	\date{}
	\maketitle

\begin{abstract}
Coded caching and device-to-device (D2D) communication are two effective techniques for alleviating network traffic. Secure transmission and file privacy have also become critical concerns in these domains. However, prevailing coded caching schemes typically assume that a user's cached content is inaccessible to others, overlooking the risk of file privacy leakage due to attacks targeting the cache itself. In this paper, we propose a secure coded caching scheme for D2D networks that guarantees both file privacy and secure delivery. We demonstrate that the proposed scheme achieves order-optimal performance when the file size is sufficiently large and the cache memory is ample.
\end{abstract}


\section{INTRODUCTION}
Mobile network data traffic continues to grow steadily, primarily driven by video streaming, which is expected to account for 76\% of all mobile data traffic by the end of 2025\cite{EMR2025}. However, the constant increase in mobile data traffic has imposed a considerable burden on the network. Caching is a highly effective solution for this problem at present. The caching system usually consists of two phases: a \emph{placement phase} during off-peak hours and a \emph{delivery phase} during peak hours. The execution of these two phases is referred to as one \emph{round}. In fact, caching reduces traffic by prefetching video content to the user's device during the placement phase, allowing requests initiated by the user during the delivery phase to be partially fulfilled locally.

To harness the full potential of caching technology, Maddah-Ali and Niesen proposed a centralized coded caching scheme in \cite{MN2014}. This scheme, which involves jointly optimizing the file segmentation, cache placement, and coded delivery phases, aims to minimise traffic during the delivery phase for all user demands. Building upon this seminal work, the coded caching paradigm has been extended to various network scenarios, including device-to-device (D2D) networks, secure transmission, and secure caching, among others. Ji, Caire, and Molisch extended the coded caching scheme to the D2D network \cite{JCM2015}. In this system, during the cache placement phase, some packets of the file are stored into the users' cache. Especially, during the delivery phase, users can directly request the remaining packets of the file from nearby cache devices, unlike the transmission model in \cite{MN2014} where they have to download from distant servers. However, beyond merely alleviating traffic load, ensuring security and privacy has become an equally critical requirement.

The primary security requirement for the delivery phase is to ensure that database files are accessible only to authorised users and not to external adversaries (i.e. eavesdroppers). This requirement has been extensively studied (e.g., see \cite{AS2015}, \cite{STC2014}, \cite{ZIY20191}). A conventional approach entails the server generating keys during the placement phase, which are then stored by users to facilitate secure transmission during the delivery phase. However, in a D2D network, the server should not know the keys used between users. Furthermore, the issue of private encoding caching has become another matter of widespread concern and is referred to as file privacy in \cite{QCN2024}. File privacy requires that during the placement phase, no user can obtain any information about files they are not authorized to access \cite{RPK2016}, \cite{RPK2017}, \cite{ZIY20192}. Indeed, if an external attacker (e.g., a curious attacker) were to obtain the complete file in this phase, the consequences could be severe, extending beyond privacy breaches \cite{DW2012}. For instance, the leaked file could be used for highly targeted phishing campaigns \cite{FHN2014}. Therefore, file privacy requires protecting files not only from unauthorised access by users, but also from external adversaries.

The existing literature commonly assumes that cached content is inaccessible during the placement phase to any entity other than the legitimate user. However, in reality, users' devices may still be subject to external attacks such as malware and local attacks, which can lead to the leakage of cached content \cite{M2024}, \cite{NZV2020}. In a caching system, a file is divided into multiple packets and distributed to all users during the placement phase.
Although attackers may be unable to retrieve all of a file's packets quickly enough to reconstruct it during this phase.
However, if cached content remains static for a long time, an attacker could gradually accumulate all its packets and eventually reconstruct the file.
If a potential leak is detected, the cached content should be proactively updated before delivery. Such an update renders any file packets obtained by an attacker beforehand obsolete, thereby preventing file reconstruction even if many packets had been accumulated. A straightforward solution would involve the remote server regenerating and redistributing the cached content to all users. However, this approach incurs substantial transmission and computation overhead. Furthermore, complete reliance on the server for reprocessing (be it keys or files) is unreliable. This creates a single point of failure, which is a significant vulnerability.

Therefore, this paper aims to address these issues by designing an encoded caching scheme that can withstand long-term attacks during the placement phase. The goal of the scheme is to minimize the total number of transmitted bits in the delivery phase while satisfying the demands of all users. To address this, we propose a \emph{proactive coded caching scheme} that ensures both file privacy and secure transmission in D2D networks.
The placement phase of this scheme consists of two stages: a \emph{prefetching stage} and a \emph{renewal stage}, responsible for the initial caching and the subsequent proactive updating of the content, respectively. In this scheme, we encode each file using a ramp secret sharing scheme \cite{BM1985} and distribute the resulting shares to the users' caches. Inspired by proactive secret sharing \cite{HJK1995}, the scheme mandates that users refresh their cached content before the delivery stage if a file leakage risk is detected. To suit the D2D caching architecture, secret keys are generated locally by leveraging the Diffie-Hellman problem (DH problem)\cite{DH1976}. Importantly, both the cache update process and the key generation are completed entirely within the placement phase. In order to reduce the complexity of key management, the scheme also incorporates key derivation function algorithm, such as the hash-based key derivation function algorithm (HKDF) \cite{KE2010}. Our analysis shows that the proposed scheme achieves an order-optimal memory-load trade-off in the regime of large file sizes and large cache memory capacities.

Paper Organization: Section \ref{sec-system} describes the system model and reviews related work. Section \ref{sec-mainresult} presents the results of the proposed proactive coded caching scheme. Section \ref{sec-example}  provides an illustrative example. Section \ref{sec-analysis} presents the performance analysis. Section \ref{sec-coclusion} concludes the paper.

\textbf{Notation}: For $n_1<n_2\in\mathbb{N}$, we denote the set $\{0,1,\ldots,n_1-1\}$ by $[n_1]$ and the set $\{n_1,n_1+1,\ldots,n_2\}$ by $[n_1,n_2]$. A vector of random variables is denoted by a bold-faced uppercase letter. The least integer greater than or equal to $x$ is ${\left\lceil{x}\right\rceil}$. Let $q$ be a prime power. Let $\mathbb{F}_q$ be the finite field of cardinality $q$, and let $\mathbb{F}_{q}^n$ be the R-dimensional vector space over $\mathbb{F}_q$.

\section{SYSTEM MODEL}
\label{sec-system}
\begin{figure}[h]
\centering
\includegraphics[width=0.4\textwidth]{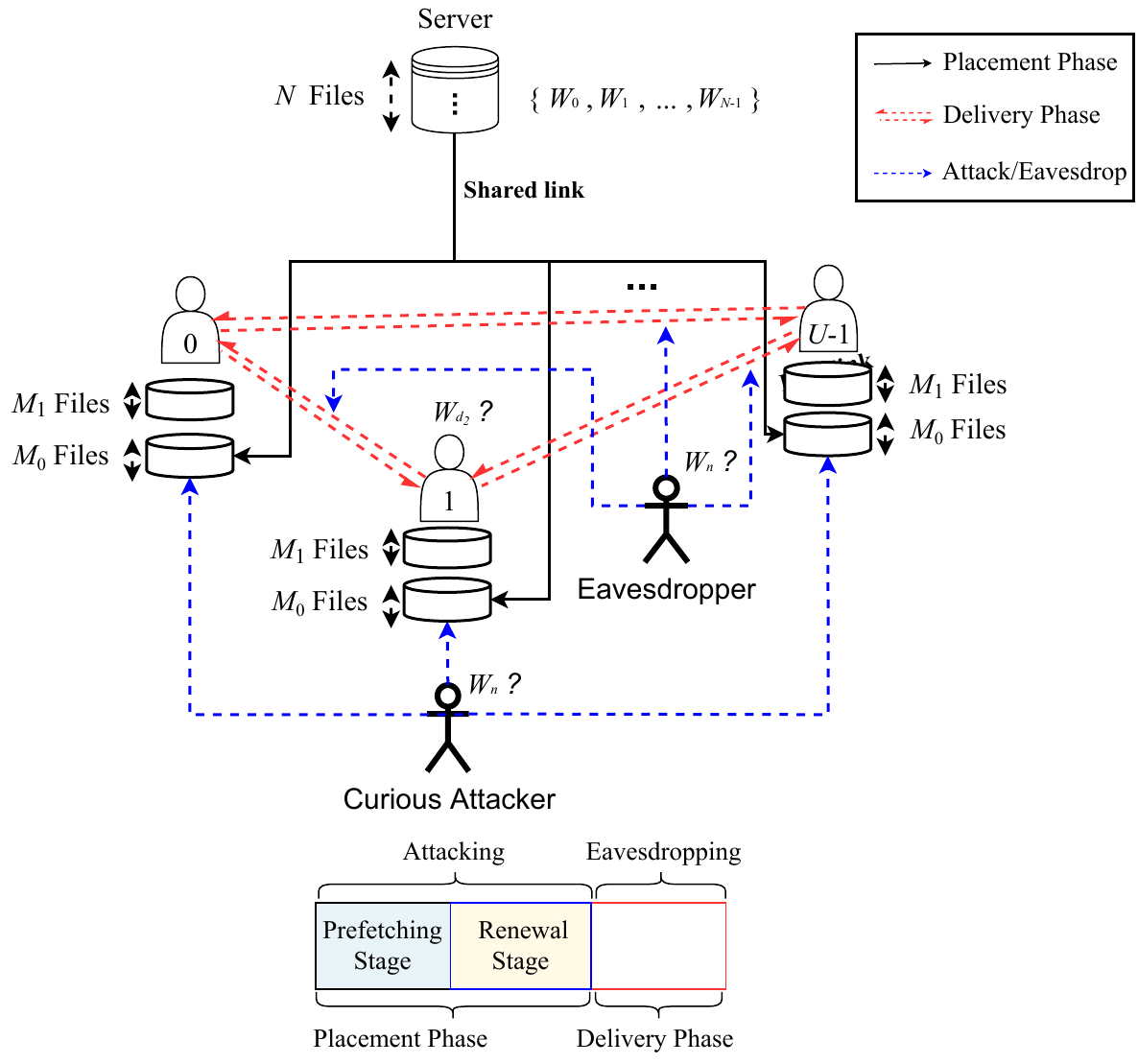}
\caption{The $(U,M,N)$ D2D proactive caching system}\label{fig-D2D-system}
\end{figure}

The D2D proactive caching system, illustrated in Fig.\ref{fig-D2D-system}, is defined as follows. A server holds a library of $N$ independent files, denoted collectively as $\mathcal{W}=\{W_0,W_1,\ldots,W_{N-1}\}$, each of size $B$ bits. The server is connected to $U$ users, denoted by $\mathcal{U}=\{0,1,\ldots,U-1\}$. We assume $B$ is sufficiently large such that each file can be partitioned into multiple subfiles of equal size. Each user is equipped with a local cache capable of storing data equivalent to $M$ entire files. Here, $M$ is represents the normalized cache size (e.g., the ratio of the cache capacity to the file size B).

The system operates in rounds, each comprising a placement phase and a delivery phase. The placement phase comprises two stages: an initial prefetching stage followed by a renewal stage during which $T$ updates are performed. Specifically, at the $t$-th update, let $Z_{0,u}^{(t)}$ denote the file cache content, and let $Z_{1,u}$ denote the content of the key cache for $u\in \mathcal{U}$, where $t\in\{0,1,\ldots,T\}$. Here, the secret key is stored in a separate cache. We assume this key cache is perfectly secure, meaning its content is never accessible to any adversary.  Let the overall cache content at user $u\in \mathcal{U}$ after the $t$-th update be denoted by $Z_{u}^{(t)}$, which comprises the disjoint file cache $Z_{0,u}^{(t)}$ and key cache $Z_{1,u}$. Their sizes are fixed at $|Z_{0,u}^{(t)}|=M_0$ files and $|Z_{1,u}|=M_1$ files, respectively, satisfying $M_0+M_1=M$ for all $t\in [T]$.

\subsection{Placement Phase}
This phase operates in two stages.
\begin{itemize}
\item \textbf{Prefetching Stage :} Each file $W_n\in\mathcal{W}$ is divided into $F$ packets of equal size, denoted as $W_n=\{W_{n,0},\ldots,W_{n,F-1}\}$. During the placement phase, each user $u\in\mathcal{U}$ independently selects and stores a subset of these packets in its local cache. This selection is made a priori, without knowledge of the specific files that will be requested later. The caching function in the beginning is denoted by
    \begin{align*}
    \phi_u^{(0)}: \mathbb{F}_2^{NB}\to \mathbb{F}_2^{M_0B}, u\in\mathcal{U}.
    \end{align*}The cached content of the user $u$ is denoted by
    \begin{align*}
    Z_{0,u}^{(0)}=\phi_u^{(0)}(\mathcal{W}).
    \end{align*}

    In addition, all users collectively generate a set of secret keys to secure the system against eavesdroppers. Let $G=\langle g \rangle$ be a multiplicative group of large prime order $p$. Each user $u\in\mathcal{U}$ selects a private random number $x_u\in \mathbb{Z}_p^*$. Let $\mathbf{x}=(x_0,x_1,\ldots,x_{U-1})$ be the vector of all user secrets. The key-generation function for user $u$ is a map $\kappa_u: (\mathbb{Z}_p^*)^U\times G \to \{0,1\}^{M_1B}$, which outputs the user's key cache content as \[Z_{1,u}=\kappa_u(\mathbf{x},g).\]
    During the extended placement phase, we consider a curious attacker who can gradually gain access to the cache content stored on user devices. Through repeated access over time, such an attacker could piece together all the packets of any given file, leading to a complete file leakage. To mitigate this risk and prevent complete file disclosure, we introduce a cache renewal mechanism between the prefetching stage and delivery phase. This mechanism involves multiple times of cache updates.
\item \textbf{Renewal Stage:} During the $t$-th update, all user generate a random vector $\mathbf{P}^{(t)}$ from a finite alphabet $\mathcal{P}$. Each user $u$ then updates its file cache using an update function
    $\psi^{(t)}_{u}:\mathcal{P}\times\mathbb{F}_2^{M_0B}\to \mathbb{F}_2^{M_0B}$. Specifically, the new cache content is computed as:
    \begin{align*}
    Z_{0,u}^{(t)}=\psi_u^{(t)}(\mathbf{P}^{(t)},Z_{0,u}^{(t-1)}).
    \end{align*}

    Intuitively, the update process essentially refreshes the cached copies of all packets belonging to a file $W_n\in\mathcal{W}$. Let $\widetilde{W}_{n,j}^{(t,u)}$ denote the cached version of packet $W_{n,j}$ for user $u$ after the $t$-th update. Then, the function $\psi^{(t)}_{u}$ updates the set $\{\widetilde{W}_{n,j}^{(t-1,u)}\}_{j=0}^{F-1}$ to produce a new set $\{\widetilde{W}_{n,j}^{(t,u)}\}_{j=0}^{F-1}$.

\end{itemize}
\subsection{Delivery Phase}
During this phase, each user $u\in\mathcal{U}$ randomly requests one file $W_{d_u}$. The demand vector of all users is denoted by $\mathbf{d}=(d_0,d_1,\ldots,d_{U-1})$. To satisfy these requests, each user $u$ broadcasts a coded message $X_{u,\mathbf{d}}$ of size $R_{u,\mathbf{d}}(M)$ files. This message is a function of the global demand $\mathbf{d}$ and the user's own cache content after the final update, $Z_{u}^{(T)}$. Formally, the transmitted signal is given by
\begin{align*}
X_{u,\mathbf{d}}=\varphi_u(\mathbf{d}, Z_{u}^{(T)}),
\end{align*}where $\varphi_u: [N]^U\times\mathbb{F}_2^{MB}\to \mathbb{F}_2^{R_{u,\mathbf{d}}(M)B}$ is the encoding function of user $u$.

The server remains silent during the delivery phase, so all user requests must be satisfied via D2D communications. Consequently, each user $u$ decodes an estimate, denoted by $\widehat{W}_{d_u}$, of its requested file $W_{d_u}$. This decoding is performed using only its own cache content $Z_{u}^{(T)}$ and the broadcast signals from all other users. Formally, the decoding function for user $u$ is

\begin{align*}
\omega_u:\mathbb{F}_2^{(\sum_{u\neq u'}{R_{u',\mathbf{d}}(M)})B}\times\mathbb{F}_2^{MB}\to \mathbb{F}_2^{B},
\end{align*}which produces the estimate as
\begin{align*}
\widehat{W}_{d_u}=\omega_u\left(\left(X_{u',\mathbf{d}}\right)_{u'\in\mathcal{U}\backslash \{u\}},
Z_{u}^{(T)}\right).
\end{align*}where $\left(X_{u',\mathbf{d}}\right)_{u' \in \mathcal{U} \setminus \{u\}}$ denotes the vector of signals received from all other users.

\subsection{System Constraints}\label{sub-SC}
The system is designed to satisfy the following constraints:
\begin{itemize}
\item \textbf{Correctness:} Each user $u\in\mathcal{U}$ should be able to decode its requested file $W_{d_u}$ with zero error probability. Equivalently,
    \begin{align}
    \label{Correctness}
        H(W_{d_u}\mid \left\{X_{u',\mathbf{d}}\right\}_{u' \in \mathcal{U} \setminus \{u\}}, \mathbf{d}, Z_{u}^{(T)})=0, ~\forall u\in\mathcal{U};
    \end{align}where $\left\{X_{u',\mathbf{d}}\right\}_{u'\in\mathcal{U}\setminus \{u\}}$ is the all received signals from users other than $u$.
\item \textbf{File Privacy:} The system ensures that no information about the files can be obtained by any party, including both users and any curious attacker, during the placement phase. On the one hand, no user should learn any information about files it did not request. Formally, for every user $u$ and for every file $W_n\neq W_{d_{u}}$,
    \begin{align}
    \label{File_Privacy_1}
        I(W_n;\left\{X_{u',\mathbf{d}}\right\}_{u' \in \mathcal{U} \setminus \{u\}},\mathbf{d},Z_{u}^{(t)})=0, ~\forall t\in[T+1].
    \end{align}

    On the other hand, an attacker who gradually collects encoded packets of a file from various users and over different update times should still not be able to fully recover the file. That is, if the collected packets correspond to all $F$ indices but come from different update time, then
    \begin{align}
    \label{File_Privacy_2}
        0\leq I\left(W_n;\{\widetilde{W}_{n,j}^{(t_j,u_{j})}\}_{j=0}^{F-1}\right)<H(W_n),
    \end{align}where $t_{0}, t_{1}, \ldots, t_{F-1}\in [T]$ are not all equal.
\item \textbf{Security:} An external eavesdropper who observes all transmitted signals during delivery phase should gain no information about any file:
    \begin{align}
    \label{Security}
        I(W_n;\left\{X_{u,\mathbf{d}}\right\}_{u \in \mathcal{U}})=0, ~\forall n\in[N].
    \end{align}
\end{itemize}

Our objective is to minimize the worst-case total transmission load incurred during the delivery phase. Formally, we define the system load $R(M)$ as  the maximum sum-rate required over all possible user demands:
\begin{align}
\label{eq-R}
R(M)=\max\limits_{\mathbf{d}\in[N]^U}\sum\limits_{u\in\mathcal{U}} R_{u,\mathbf{d}}(M).
\end{align}
We seek to achieve this minimization by designing a proactive coded caching scheme that adheres to the correctness, privacy, and security constraints specified in the previous content. Correspondingly, we introduce the following definition.

\begin{definition}[{Proactive coded caching scheme}]\rm
\label{defi-MR}
A memory-load pair $(M,R)$ is said to be \emph{achievable} if, for cache size $M$, there  exists a scheme satisfying all constraints \eqref{Correctness}-\eqref{Security} and achieving a transmission load $R$. Such a scheme is called a proactive coded caching scheme. The optimal load-memory function is defined as
\begin{align*}
R^*(M)=\inf\{R:(M,R) \text{~is a achievable}\}.
\end{align*}
\end{definition}

We now present some auxiliary definitions and known results that will be used in our analysis.

\begin{definition}[{DH problem \cite{DH1976}}] \rm ~Let $G=<g>$ be a finite cyclic group. The problem of computing $g^{x_0x_1}$ from $g^{x_0}$ and $g^{x_1}$ is called the Diffie-Hellamn problem, where $x_0$ and $x_1$ are chosen independently and uniformly in $\mathbb{Z}_{|G|}^*$.
\end{definition}

Solving the DH problem is equivalent to solving the discrete logarithm problem, as it requires extracting the secret exponent $x_0$ from its public form $g^{x_0}$.

\begin{definition}[{Ramp secret sharing scheme \cite{DLW2021},\cite{M2018}}]\rm
\label{defi-ramp}
 For an $(r,m,k)$ ramp secret sharing scheme, where $r$, $m$ and $k$ be positive integers such that $r\leq m\leq k$. Let $\mathcal{A}$ be a general $|\mathcal{A}|$-ary symbols and $\mathbf{w}\in \mathcal{A}^{m-r}$ is a secret vector. The scheme consists of the following algorithm:
\begin{itemize}
\item $\mathsf{Distribution ~ algorithm}$: It is a map
\begin{align*}
D:~ \mathcal{A}^{m-r}\times \mathcal{R} \mapsto \mathcal{A}^{k},
\end{align*}i.e., it maps $m-r$ secrets from $\mathcal{A}$ and a random string from $\mathcal{R}$ to $k$ shares $s_i\in\mathcal{A}$, $i\in[k]$.

\item $\mathsf{Reconstruction ~ algorithm}$: It satisfies the following properties.
\begin{itemize}
\item \textbf{Correctness:} For any $A\subseteq[k]$ of size $|A|\geq m$, the reconstruction algorithm always reveals the $m-r$ secrets $\mathbf{w}$ with shares $s_A$, i.e., $H(\mathbf{w};s_A)=0$. The $m$ is called reconstruction threshold.
\item \textbf{Privacy:} For any $A\subseteq[k]$ of size $|A|\leq r$, the privacy threshold $r$ ensures no information about the secret vector can be learned with shares $s_A$, i.e., $I(\mathbf{w};s_A)=0.$
\item \textbf{Ramp Security:} If $r<m$ and $A$ is a subset of $[n]$ such that $r<|A|<m$, then the secrets $\mathbf{w}$ cannot be reconstructed with shares $s_A$, but partial information is revealed, i.e., $0<I(\mathbf{w};s_A)<H(\mathbf{w}).$
\end{itemize}
\end{itemize}
\end{definition}

In particular, for the $(r,m,k)$ ramp secret sharing scheme introduced above, the optimal size of each share is $\frac{H(\mathbf{w})}{m-r}$. In our system, the server utilizes this scheme to protect each file $W_n\in\mathcal{W}$. Specifically, for each file, the server generate $k$ random vectors called shares. The scheme guarantees that: (i) the file can be reconstructed from any set of $m$ or more shares, and (ii) any set of $r$ or fewer shares reveals no information about the file. 

The following example illustrates the share generation process:

\begin{example}\rm
Let file $W_0$ be $8$ bits long, with parameters $(r,m,k)=(2,5,7)$. Suppose $W_0=(1,1,0,0,1,0,1,0)$. Since we need to divide it into $m-r=3$ subpackets, we first append one zero bit to make its length $9$, resulting in $W_0'=\{W_{0,0},W_{0,1},W_{0,2}\}$. This vector is then partitioned into three $3$-bit blocks: \[W_{0,0}=(1,1,0), W_{0,1}=(0,1,0), W_{0,2}=(1,0,0).\] We work over the finite field $\mathbb{F}_{2^{3}}=\{0,1,\alpha, \ldots, \alpha^6\}$. Each $3$-bit block corresponds to a field element: \[(1,1,0)\mapsto \alpha^5, ~(0,1,0)\mapsto \alpha, ~(1,0,0)\mapsto \alpha^5.\] Thus, the file is represented as $W_0'=(\alpha^5, \alpha, \alpha^3)\in (\mathbb{F}_{2^3})^3$.

Now, consider a polynomial $f(x)=a_0+a_1x+a_2x^2+a_3x^3+a_4x^4$ of degree $4$, where we set the coefficients $a_2=\alpha^5$, $a_3=\alpha$, and $a_4=\alpha^3$. The remaining coefficients $a_0$ and $a_1$ are chosen uniformly at random from $\mathbb{F}_{2^3}$. To generate the $7$ shares, we evaluate $f(x)$ at $7$ distinct non-zero field elements, e.g., at $x=1,2,\ldots,7$. Each evaluation constitutes one share. Since $f(x)$ has degree $4$, any $5$ out of these $7$ shares are sufficient to reconstruct the polynomial via Lagrange interpolation.  Conversely, any $2$ or fewer shares reveal no information about the file.
\end{example}

\section{MAIN RESULTS}
\label{sec-mainresult}
In this section, we present our results on the achievable memory-load tradeoff and the corresponding proactive coded caching scheme. The proof of achievability is analyzed in Subsection \ref{sec-Achievability}.

\begin{theorem}\rm
\label{theorem-1}
For an $(U,M,N)$ D2D proactive caching system described in Section \ref{sec-system}, there exists a scheme achieving the following memory-load pairs:
\begin{align*}
(M,R)=\left(\frac{Nl}{U-l}+\frac{1}{l}, \frac{2U(N+M)}{MU+1+\sqrt{(MU-1)^2-4NU}}\right),
\end{align*}
where $l\in[1,U-1]$.
\end{theorem}

The remainder of this section is devoted to proving Theorem \ref{theorem-1} by constructing an explicit proactive coded caching scheme that achieves the above trade-off points.

\subsection{Placement Phase}
\subsubsection{Prefetching Stage}
In the initial prefetching stage ($t=0$), the server encodes each file using an $\left(l{U-1 \choose l-1}, l{U \choose l}, l{U \choose l}\right)-$ ramp secret sharing scheme. Each file $W_n\in\mathcal{W}$ is first divided into $m-r$ equal size packets, denoted as \[W_n=\{W_{n,0},W_{n,1},\ldots,W_{n,m-r-1}\}.\] This requires the file size $B$ to be such that $L=\frac{B}{m-r}=\frac{B}{l{U-1 \choose l}}$ is a positive integer. For encoding, the server generates for each file $W_n$ a random polynomial of degree $m-1$:
\begin{align*}
f_n^{(0)}(x)=\sum_{i=0}^{m-1}a^{(0)}_{n,i}x^i\in \mathbb{F}_{2^L}[x].
\end{align*}

We set $a^{(0)}_{n,j+r}=W_{n,j}$, where $j\in [m-r]$. The server then evaluates $f_n^{(0)}(x)$ at $k$ pre-defined public points to generate the shares. Specifically, for every subset $\mathcal{T}\subset\mathcal{U}$ of size $|\mathcal{T}|=l$ and for each $i=0,1,\ldots,l-1$, there is a corresponding public evaluation point $c_\mathcal{T}^i\in \mathbb{F}_{2^L}$. The share for file $W_n$ associated with $(\mathcal{T},i)$ is computed as
\begin{align*}
S_{n,\mathcal{T}}^{i,(0)}=f_n^{(0)}(c_\mathcal{T}^i)\in \mathbb{F}_{2^L}.
\end{align*}

Each share has a size of $F_s=L$ bits. Based on this encoding, the cache content and its size for each user $u\in\mathcal{U}$ are determined as follows:
\begin{align}
\label{eq-M_0}
\begin{aligned}
&{Z}_{0,u}^{(0)}=\{S_{n,\mathcal{T}}^{i,(0)}\mid u\in\mathcal{T}, \mathcal{T}\subset\mathcal{U}, |\mathcal{T}|=l, n\in[N], i\in[l]\},\\
&\left|{Z}_{0,u}^{(0)}\right|=N\times l\times {U-1 \choose l-1}\times F_s=\frac{Nl}{U-l}B=M_0B.
\end{aligned}
\end{align}

Furthermore, to ensure secure communication during the delivery phase, users must establish a set of common secret keys. Specifically, in our scheme, every user subset $\mathcal{L}\subseteq\mathcal{U}$ of size $l+1$ possesses a shared key. This key, agreed upon by all members of set $\mathfrak{L}$, is referred to as the Master Secret Key (MSK) for that subset and is denoted $K_{\mathcal{L}}\in \mathbb{F}_{2^L}$. We denote the collection of all such subsets of size $l+1$ by $\mathfrak{L}$, and the collection of all subsets of size $l$ by $\mathfrak{T}$.

A direct implementation would have each subset $\mathcal{L}\in\mathfrak{L}$ independently run a Group Key Agreement (GKA) protocol \cite{ITDW1982}. However, this  is highly inefficient because user membership overlaps extensively: each user $u$ belongs to ${U-1 \choose l}$ different subsets in $\mathfrak{L}$. Consequently, a user would need to perform essentially the same expensive operations many times for different but overlapping groups. To overcome this inefficiency, we propose a coordinated key establishment scheme that treats the entire user set as a single group. By leveraging broadcast communication and the hardness of the Diffie-Hellman problem, our scheme allows all MSKs $\{K_{\mathcal{L}}|\mathcal{L}\in\mathfrak{L}\}$ to be derived simultaneously with minimal per-user computation, eliminating the redundancy inherent in the naive approach. The complete procedure is outlined in Algorithm \ref{alg-upkeyd2d-ppk}.

\begin{breakablealgorithm}
\caption{The generation of the Master Secret Key for groups of every $l+1$ based on DH problem}\label{alg-upkeyd2d-ppk}
\begin{algorithmic}[1]
\Procedure {Setup} {$\mathcal{U}$, $l$, $L$}
\State Let $G=<g>$ be a multiplicative group of prime order $p$, and let $p$ be a large prime of length $L$ bits.
\State For user $u\in \mathcal{U}$ selects a random number $x_u\in \mathbb{Z}_p^{*}$. Each user computes $g^{x_u}$ and broadcasts it.
\For{$l'=1,2,\ldots, l$}
\State $\mathfrak{T'} \leftarrow \{\mathcal{T'}|\mathcal{T'}\subset \mathcal{U}, |\mathcal{T'}|=l'\}$
\State Each user raises the received intermediate key value to the power of its own exponent. Each User $u\in \mathcal{U}$ computes $K_{\{\mathcal{T}'\cup u\}}=\left(g^{\prod\{{x_{u'}}|u'\in \mathcal{T'}, u\notin \mathcal{T'}, \mathcal{T'}\in \mathfrak{T}\}}\right)^{x_u}$. It is estimated that each user should calculate approximately $\lceil{{U\choose l'}/{U}}\rceil$ values.
\If {$l'<l$}
\State Each User $u\in \mathcal{U}$ broadcasts the results $K_{\{\mathcal{T}'\cup u\}}$ to all user.
\Else
\State Every user group $\mathcal{L}\in\mathfrak{L}$ possesses a common key $K_{\mathcal{L}}$.
\EndIf
\EndFor
\EndProcedure
\end{algorithmic}
\end{breakablealgorithm}

\begin{remark}\rm
For each subset $\mathcal{T}'\subset\mathcal{U}$, where $|\mathcal{T}'|=l'$, we designate a specific user in $\mathcal{T}'$ to handle computations related to $\mathcal{T}'$. Let $\mathcal{T}'=\{u_0,u_1,\ldots,u_{l-1}\}$ with users indexed in increasing order.  The designated user index is chosen as $f(\mathcal{T})=\sum_{i=1}^{t}{u_i}\mod {l'}$. Thus, user $u_{f(\mathcal{T})}\in\mathcal{T}'$ is responsible for $\mathcal{T}'$. Under this scheme, each user is assigned to approximately $\lceil{{U\choose l'}/{U}}\rceil$ subsets, balancing the computational load.
\end{remark}

During the delivery phase, each user group $\mathcal{L}\in\mathfrak{L}$ may require multiple session keys. To simplify key management, we employ an HKDF (HMAC-based Key Derivation Function) \cite{KE2010}. All keys needed for a group $\mathcal{L}$ can be derived from its MSK $K_{\mathcal{L}}$. As shown in \cite{KE2010}, public values generated during the MSK agreement phase can be used as salt to strengthen the HKDF. Once users agree on a common hash function, they can use the same MSK $K_{\mathcal{L}}$ and public parameters to generate identical derived keys. Consequently, each user only needs to store the MSKs of the groups it belongs to, rather than all derived session keys.

Consequently, the following content is cached by the user $u\in \mathcal{U}$, namely,
\begin{align}
\label{eq-M_1}
\begin{aligned}
{Z}_{1,u}=\{K_{\mathcal{L}} \left|\right. u\in\mathcal{L}, \mathcal{L}\in\mathfrak{L}\}.
\end{aligned}
\end{align}Its size in bits is given by
\begin{align}
\label{eq-M_1}
\begin{aligned}
\left|{Z}_{1,u}\right|={U-1 \choose l}\times F_s=\frac{B}{l}=M_1B ~\text{(bits)}.
\end{aligned}
\end{align}

Consequently, during period 0, the total cache content of user $u$ is the union of its file cache and key cache:
\begin{align*}
Z_{u}^{(0)}={Z}_{0,u}^{(0)}\cup{Z}_{1,u}.
\end{align*}

Thus, by Equations \eqref{eq-M_0} and \eqref{eq-M_1}, the storage for each user is

\begin{align*}
M_0B+M_1B=\left(\frac{Nl}{U-l}+\frac{1}{l}\right)B~ \text{(in bits)},
\end{align*}i.e., $M=\frac{Nl}{U-l}+\frac{1}{l}$ files. It is observed that the cache capacity constraints for all users are satisfied.

\begin{remark}\rm
For $M=\frac{Nl}{U-l}+\frac{1}{l}$, we can get
\begin{align}
\label{eq-l}
l=\frac{MU+1+\sqrt{(MU-1)^2-4NU}}{2(N+M)}.
\end{align}
\end{remark}
\subsubsection{Renewal Stage}
The system performs periodic updates to refresh the cached shares. Here, we assume that the set of files to be updated is $\mathcal{A}_t\subseteq\mathcal{W}$. In the $t$-th update, $t=1,2\ldots,T$, the following procedure is executed for each file $W_n\in\mathcal{A}_t$:

i) Polynomial Generation: Each user $u\in\mathcal{U}$ independently generates a random polynomial of degree $r-1$: \[h_{n,u}^{(t)}(x)=\sum_{i=0}^{r-1}b^{u,(t)}_{n,i}x^i\in \mathbb{F}_{2^L}[x],\] where the coefficients are chosen uniformly at random from $\mathbb{F}_{2^L}$. For each user group $\mathcal{T}\in\mathfrak{T}$ that $u$ belongs to, user $u$ computes the values $h_{n,u}^{(t)}(c_\mathcal{T}^i)$, where $c_\mathcal{T}^i\in \mathbb{F}_{2^L}$ is a public evaluation point associated with group $(\mathcal{T},i)$ and $i=0,1,\ldots,l-1$.

ii) Secure Exchange: For each subset $\mathcal{T}\in\mathfrak{T}$, each user $u\in\mathcal{T}$ involved need to perform the following operations. To securely broadcast this value to all other members of $\mathcal{T}$, user $u$ encrypts it using a temporary session key $EK_{\mathcal{T}}^{i}\in \mathbb{F}_{2^L}$. This session key is derived from the group's Master Secret Key $K_{\mathcal{T}}$ (e.g., using HKDF with a specific salt). User $u$ broadcasts signals $\mathcal{X}_{n,u}$ and $\mathcal{Y}_{n,u}$ as follows:
\begin{align}
\label{eq-XY}
\begin{aligned}
\mathcal{X}^{(t)}_{n,u}&=\{h_{n,u}^{(t)}(c_{\mathcal{T}}^i)\oplus EK_{\mathcal{T}}^{i}|u\in \mathcal{T}, i\in[l]\},\\
\mathcal{Y}^{(t)}_{n,u}&=\{h_{n,u}^{(t)}(c_{\mathcal{T}}^i)\oplus K_{\mathcal{L}}^{i}|u\notin\mathcal{T},\mathcal{L}=\mathcal{T}\cup u, i\in[l+1]\};
\end{aligned}
\end{align}where $\{K_{\mathcal{L}}^{i}\}_{i=0,1,\ldots,l}$ is derived from $K_{\mathcal{L}}$ and public values by the HKDF algorithm. The public values are generated from $EK_{\mathcal{T}}^{i}\in \mathbb{F}_{2^L}$, $i=0,1,\ldots,l$. It is evident that, in the case of $l = 1$, the requirement for each user is only to broadcast $\mathcal{Y}_{n,u}$.

iii) Share Refresh: Using the broadcast signals $ \{\mathcal{X}^{(t)}_{n,u'},\mathcal{Y}^{(t)}_{n,u'}\mid u'\in \mathcal{U}\setminus\{u\}\}$ and the public evaluation points $\{c_{\mathcal{T}}^i\mid{\mathcal{T}\in \mathfrak{T},i\in[l]}\}$, each user $u$ first decodes the relevant polynomial evaluations as follows:
\begin{align*}
\begin{aligned}
&\mathcal{X}^{(t)}_{n,u'}\xrightarrow{\text{decode}}\{h_{n,u'}^{(t)}(c_{\mathcal{T}}^i)|\{u,u'\}\subset\mathcal{T},\mathcal{T}\in\mathfrak{T},i\in[l]\},\\
&\mathcal{Y}^{(t)}_{n,u'}\xrightarrow{\text{decode}}\{h_{n,u'}^{(t)}(c_{\mathcal{T}}^i)|u\in\mathcal{T}, \mathcal{T}\in\mathfrak{T},i\in[l]\}.
\end{aligned}
\end{align*}

After collecting all such evaluations $\{h_{n,u}^{(t)}( c^i_{\mathcal{T}})\}_{u\in\mathcal{U}}$ for $(\mathcal{T},i)$, each user $k\in\mathcal{T}$ computes the aggregate value \[h_n^{(t)}(c^i_{\mathcal{T}})=\sum_{u\in\mathcal{U}} h_{n,u}^{(t)}( c^i_{\mathcal{T}})\in\mathbb{F}_{2^L}.\] This aggregate value is then used to refresh the corresponding cached share. The updated share for file $W_n$ at point $c^i_{\mathcal{T}}$ becomes:
\begin{align}
\label{eq-update}
~S_{n,\mathcal{T}}^{i,(t)}=S_{n,\mathcal{T}}^{i,(t-1)}+h_n^{(t)}(c_{\mathcal{T}}^i)\in\mathbb{F}_{2^L},
\end{align}which replaces the old share in the cache $Z_{0,k}^{(t-1)}$ to form the new cache state $Z_{0,k}^{(t-1)}$. Thus, from the broadcast signals, each user $k$ obtains a total of ${U-1 \choose l-1}$ polynomial evaluations.

This update operation effectively modifies the underlying polynomial that generates the shares. Let the polynomial after the $(t-1)$-th update be \[f_n^{t-1}(x)=\sum_{i=0}^{m-1}a^{(t-1)}_{n,i}x^i,\] and let the aggregate user polynomial be \[h_n^{(t)}(x)=\sum_{i=0}^{r-1}b^{(t)}_{n,i}x^i.\] Then the new polynomial after the $t$-th update is
\begin{align}
\label{eq-ft}
\begin{split}
f_n^{(t)}(x)&=f_n^{(t-1)}(x)+h_n^{(t)}(x)\\
&=\sum_{i=0}^{r-1}(a^{(t-1)}_{n,i}+b^{(t)}_{n,i})x^i+\sum_{i=r}^{m-1}a^{(t-1)}_{n,i}x^{i}.
\end{split}\end{align}Note that the coefficients $a^{(t-1)}_{n,i}$ for $i\geq r$ correspond to the original file packets $W_{n,i-r}$ and remain unchanged throughout the renewal stage.

In summary, after the $t$-th update, the shares for file $W_n$ are regenerated by the new polynomial $f^{(t)}_{n}(x)$. The set of updated shares for $W_n$ is
\begin{align*}
\{S_{n,\mathcal{T}}^{i,(t)}\mid\mathcal{T}\in\mathfrak{T},i\in[l]\}.
\end{align*}We can assume that the shares of the other files $W_{n'}\not\in{\mathcal{A}_t}$ were updated in the same way for $t-1$ times. Consequently, the stored content of the user $u\in\mathcal{U}$ is refreshed as
\begin{align*}
Z_{0,u}^{(t)}=\{S_{n,\mathcal{T}}^{i,(t)},S_{n',\mathcal{T}}^{i,(t-1)}\mid\mathcal{T}\in\mathfrak{T}, u\in \mathcal{T},i\in[l],n\neq n',W_n\in\mathcal{A}_t\}.
\end{align*}This completes the description of the renewal stage.
\subsection{Delivery Phase}
During the delivery phase, we consider the worst-case scenario where all users request distinct files. Let user $u\in\mathcal{U}$ request the file $W_{d_u}\in\mathcal{W}$, and denote the demand vector by $\mathbf{d}=(d _0,d_1,\ldots,d_{U-1})$. To satisfy these demands, each user $u$ broadcasts a coded signal $X_{u,\mathbf{d}}$. The signal is described below. For each subset $\mathcal{L}\in\mathfrak{L}$ that contains $u$, user $u$ generates one transmitted symbol. Let the users in $\mathcal{L}$ be indexed in increasing order. User $u$ transmits the following signals,
\begin{align}
\label{eq-delivery}
X_{u,\mathbf{d}}=\{K_{\mathcal{L}}^{j}
\bigoplus\limits_{k\in\mathcal{L}\cap{A_t}}S_{d_{k},\mathcal{L}\backslash \{k\}}^{i,{(t)}}
\bigoplus\limits_{k'\in\mathcal{L}\setminus A_t}S_{d_{k'},\mathcal{L}\backslash \{k'\}}^{i,{(t-1)}}\mid u\in\mathcal{L}, \mathcal{L}\in\mathfrak{L}\},
\end{align}where $\{K_{\mathcal{L}}^{j}\}_{j\in[l+1,2l+1]\}}$ are a one-time keys derived from MSK $K_{\mathcal{L}}$, with index $j$ chosen uniquely for each $\mathcal{L}$; $A_t=\{k\mid k\in\mathcal{U},W_{d_k}\in\mathcal{A}_t\}$ denotes the set of users requesting the file at the $t$-th update.

Therefore, using Equation \eqref{eq-R}, the total number of the transmission load is given by
\begin{align}
\label{eq-Rs}
\begin{split}
R(M)=\sum\limits_{u\in\mathcal{U}} R_{u,\mathbf{d}}(M)&=U\times\frac{\left[{U-1 \choose l}\times F_s\right]}{B}\\
&=U{U-1 \choose l}\times \frac{1}{l{U-1 \choose l}}\\
&=\frac{U}{l}.
\end{split}
\end{align}By Equations \eqref{eq-l} and \eqref{eq-Rs}, we can get
\begin{align*}
R(M)=\frac{2U(N+M)}{MU+1+\sqrt{(MU-1)^2-4NU}}.
\end{align*}

\subsection{Achievability}
\label{sec-Achievability}
In this subsection, we prove Theorem \ref{theorem-1} by demonstrating that the proposed scheme satisfies the constraints described in Section \ref{sub-SC}.

\textbf{Correctness:} During the delivery phase, each user $u \in \mathcal{U}$ requests a file and receives coded transmissions from every other user $u' \neq u$. For any subset $\mathcal{L} \in \mathfrak{L}$ that contains both $u$ and $u'$, the coded transmissions received by user $u$ can be written as
\begin{itemize}
\item If $W_{d_u}\in\mathcal{A}_t$, i.e., user $u$ requests the $t$-th updated shares of the file. We have
\begin{align}
&K_{\mathcal{L}}^{j}\oplus S_{d_{u},\mathcal{L}\backslash \{u\}}^{i,{(t)}}
\bigoplus\limits_{\substack{k'\in{A_t}\cap B\\B=\mathcal{L}\setminus \{u',u\}}}S_{d_{k'},\mathcal{L}\backslash \{k'\}}^{i,{(t)}}
\bigoplus\limits_{\substack{k'\in B\setminus A_t\\B=\mathcal{L}\setminus \{u'\}}}S_{d_{k'},\mathcal{L}\backslash \{k'\}}^{i,{(t-1)}};
\end{align}
\item  If $W_{d_u}\notin\mathcal{A}_t$, i.e., user $u$ does not request shares from the $t$-th update of the file. We have
\begin{align}
&K_{\mathcal{L}}^{j}\oplus S_{d_{u},\mathcal{L}\backslash \{u\}}^{i,{(t-1)}}
\bigoplus\limits_{\substack{k'\in{A_t}\cap B\\B=\mathcal{L}\setminus \{u'\}}}S_{d_{k'},\mathcal{L}\backslash \{k'\}}^{i,{(t)}}
\bigoplus\limits_{\substack{k'\in B\setminus A_t\\B=\mathcal{L}\setminus \{u',u\}}}S_{d_{u},\mathcal{L}\backslash \{u\}}^{i,{(t-1)}};
\end{align}
\end{itemize}where index $j\in[l+1,2l+1]$ with it chosen uniquely for each $\mathcal{L}$; $A_t=\{k\mid k\in\mathcal{U},W_{d_k}\in\mathcal{A}_t\}$ denotes the set of users requesting the file at the $t$-th update.
User $u$ can obtain
\begin{align}
\bigoplus\limits_{\substack{k'\in{A_t}\cap B\\B=\mathcal{L}\setminus \{u',u\}}}S_{d_{k'},\mathcal{L}\backslash \{k'\}}^{i,{(t)}}
\bigoplus\limits_{\substack{k'\in B\setminus A_t\\B=\mathcal{L}\setminus \{u'\}}}S_{d_{k'},\mathcal{L}\backslash \{k'\}}^{i,{(t-1)}} ~~\text{or} \bigoplus\limits_{\substack{k'\in{A_t}\cap B\\B=\mathcal{L}\setminus \{u'\}}}S_{d_{k'},\mathcal{L}\backslash \{k'\}}^{i,{(t)}}
\bigoplus\limits_{\substack{k'\in B\setminus A_t\\B=\mathcal{L}\setminus \{u',u\}}}S_{d_{u},\mathcal{L}\backslash \{u\}}^{i,{(t-1)}},
\end{align}from the content of its cache $Z_{0,u}$. We now consider the term $K_{\mathcal{L}}^{j}$. Using the MSK $K_{\mathcal{L}}$ stored in its secret key cache $Z_{1,u}$, user $u$ can derive the session key $K_{\mathcal{L}}^{j}$. From these transmissions, user $u$ can then decode share of its requested file:
\begin{align}
S_{d_{u},\mathcal{L}\backslash \{u\}}^{i,{(t)}} ~~\text{or}~~
S_{d_{u},\mathcal{L}\backslash \{u\}}^{i,{(t-1)}}.
\end{align}

Hence, user $u$ can obtain $(U-1){U-2 \choose l-1}=l{U-1 \choose l}=m-r$ distinct shares of its requested file $W_{d_u}$. In addition, user $u$ already stores $r$ shares of $W_{d_u}$ in its cache during the renewal stage. Therefore, user $u$ collectively possesses $m$ different shares of $W_{d_u}$ by the same update time. Without loss of generality, we assume that file $W_{d_u}$ have thought $T$-th update shares. We have
\begin{align}
H(W_{d_u};\underbrace{\{S^{i,(T)}_{d_u,\mathcal{L}\setminus \{u\}}\}_{\substack{\mathcal{L}\in\mathfrak{L},\\i\in[l]}}}_{m-r},\underbrace{\{S^{i,(T)}_{d_u,\mathcal{T}}\}_{ \substack{u\in\mathcal{T}, i\in[l],\\\mathcal{T}\in\mathfrak{T}}}}_{r})=0.
\end{align} By the reconstruction property of the $(r,m,k)$-ramp secret sharing scheme used in the prefetching stage, any $m$ shares uniquely determine the original file, provided they belong to the same update time. Since user $u$ has now obtained $m$ such shares, it can reconstruct its requested file $W_{d_u}$.

\textbf{File Privacy:} The file privacy stems from the underlying $(r,m,k)$-ramp secret sharing scheme and the renewal mechanism. We analyze it from two aspects:

i) Privacy against Users: The files are independently and uniformly distributed. Consider a user $u$ who has not requested file $W_{n}\in\mathcal{W}\setminus\{W_{d_u}\}$. We aim to show that user $u$ cannot reconstruct file $W_n$. Under our prefetching and renewal stages, the number of shares of $W_n$ stored in user $u$'s cache does not exceed the privacy threshold $r$. In the renewal stage, assume that the shares of file $W_n$ has been updated $t$ times. By Equation \eqref{eq-ft}, user $u$ can obtain new shares of file $W_n$ as follows:
\begin{align}
\underbrace{f_n^{(t)}(c^{i}_{\mathcal{T}})}_{\underset{\text{New shares}}{\underline{\{S^{i,(t)}_{n,\mathcal{T}}\}_{ \substack{u\in\mathcal{T},i\in[l]}}}}}&=\underbrace{f_n^{(t-1)}(c^{i}_{\mathcal{T}})}_{\underset{\text{Old shares}}{\underline{\{S^{i,(t-1)}_{n,\mathcal{T}}\}_{ \substack{u\in\mathcal{T},i\in[l]}}}}}+h_n^{(t)}(c^{i}_{\mathcal{T}}),
\end{align}From the above equation, we can see that the number of shares of file $W_n$ held by user $u$ remains unchanged, and these shares are always generated by evaluating the corresponding polynomial at the fixed points $\{c^{i}_{\mathcal{T}}\mid \in[l], \mathcal{T}\in\mathfrak{T}\}$ in each update. We have
\begin{align}
\label{eq-pau1}
I(W_n;\left\{X_{u',\mathbf{d}}\right\}_{u' \in \mathcal{U} \setminus \{u\}},\mathbf{d},\{S_{n,\mathcal{T}}^{i,{(t)}}\}_{\substack{u\in\mathcal{T},\\\mathcal{T}\in\mathfrak{T}}})
&=\underbrace{I(W_n;\{S_{n,\mathcal{T}}^{i,{(t)}}\}_{\substack{u\in\mathcal{T},\\\mathcal{T}\in\mathfrak{T}}})}_{r} +I(W_n;\left\{X_{u',\mathbf{d}}\right\}_{u' \in \mathcal{U} \setminus \{u\}},\mathbf{d} \mid \{S_{n,\mathcal{T}}^{i,{(t)}}\}_{\substack{u\in\mathcal{T},\\\mathcal{T}\in\mathfrak{T}}})\\
&\stackrel{(a)}{=}I(W_n;\left\{X_{u',\mathbf{d}}\right\}_{u' \in \mathcal{U} \setminus \{u\}},\mathbf{d} \mid \{S_{n,\mathcal{T}}^{i,{(t)}}\}_{\substack{u\in\mathcal{T},\\\mathcal{T}\in\mathfrak{T}}})\label{eq-agu2}, ~\forall t\in[T+1].
\end{align}Here, $(a)$ since the ramp scheme property, any set of at most $r$ shares reveals zero information about $W_n$.

If the file $W_n$ is not requested during the delivery phase, then the term $\left\{X_{u',\mathbf{d}}\right\}_{u' \in \mathcal{U} \setminus \{u\}}$ in Equation \eqref{eq-agu2} does not include any information about file $W_n$. Hence, we have $I(W_n;\left\{X_{u',\mathbf{d}}\right\}_{u' \in \mathcal{U} \setminus \{u\}},\mathbf{d},\{S_{n,\mathcal{T}}^{i,{(t)}}\}_{\substack{u\in\mathcal{T},\\\mathcal{T}\in\mathfrak{T}}})=0$. Next, consider the case where $W_{d_{v}}=W_{n}$, i.e., user $v$ request file $W_n$. From Equation \eqref{eq-delivery}, we see that user $u$ can obtain the following information about $W_n$:
\begin{align}
X_{u',\mathbf{d}}\rightarrow&\{\underbrace{S_{n,\mathcal{L}\backslash \{v\}}^{i,{(t)}}}_{\text{User $u$ cache}}\oplus K_{\mathcal{L}}^{j}
\bigoplus\limits_{k\neq v\in\mathcal{L}\cap{A_t}}S_{d_{k},\mathcal{L}\backslash \{k\}}^{i,{(t)}}
\bigoplus\limits_{k'\in\mathcal{L}\setminus A_t}S_{d_{k'},\mathcal{L}\backslash \{k'\}}^{i,{(t-1)}}\mid \mathcal{L}\in\mathfrak{L}, \{u,u',v\}\subseteq\mathcal{L}\} ~~\text{and}\\
X_{u',\mathbf{d}}\rightarrow&\{S_{n,\mathcal{L}\backslash \{v\}}^{i,{(t)}}
\oplus \underbrace{{K_{\mathcal{L}}^{j}}
\bigoplus\limits_{k\neq v\in\mathcal{L}\cap{A_t}}S_{d_{k},\mathcal{L}\backslash \{k\}}^{i,{(t)}}
\bigoplus\limits_{k'\in\mathcal{L}\setminus A_t}S_{d_{k'},\mathcal{L}\backslash \{k'\}}^{i,{(t-1)}}}_{\text{User $u$ not cache}}\mid u\notin\mathcal{L}, \mathcal{L}\in\mathfrak{L}, \{v, u'\}\subseteq\mathcal{L}\}
\end{align}From the above equations, user $u$ gains no information about the file $W_n$ it did not request; that is, \[I(W_n;\left\{X_{u',\mathbf{d}}\right\}_{u' \in \mathcal{U} \setminus \{u\}},\mathbf{d},\{S_{n,\mathcal{T}}^{i,{(t)}}\}_{\substack{u\in\mathcal{T},\\\mathcal{T}\in\mathfrak{T}}})=0.\] This implies that no user can obtain any information about unrequested files from its cache.

ii) Privacy against Curious Attacker: An attacker who eavesdrops over time may obtain shares from multiple update times, collected in a set $\mathcal{T}'=\{t_j: j\in[F]\}$ where each $t_j\in[T+1]$ denotes an update index. Crucially, shares from different update times are generated using independent random polynomials $\{f^{(t')}_{n}(x)\mid t'\in\mathcal{T}'\}$. Due to the properties of the $(r,m,k)$-ramp secret sharing scheme and operation in the renewal stage, even if the attacker collects more than $r=l\binom{U-1}{l-1}$ shares, these shares do not correspond to a single encoding polynomial and thus cannot be combined to reconstruct the file. Assume file $W_n$ has updated $t$ times, and each share is associated with an update index $t_{(i,\mathcal{T})}\in[t+1]$. The attacker gradually collects these shares over different update times. we have
\begin{align}
I\left(W_n;\{S^{i,(t_{(i,\mathcal{T})})}_{n,\mathcal{T}}\}_{\mathcal{T}\in\mathfrak{T},i\in[l]}\right)
\stackrel{(a)}{<}I\left(W_n;\{S^{i,(t)}_{n,\mathcal{T}}\}_{\mathcal{T}\in\mathfrak{T},i\in[l]}\right)=H(W_n),
\end{align}where $\{t_{(i,\mathcal{T})}\}_{j\in[l],\mathcal{T}\in\mathfrak{T}}$ are not all equal.
Here, $(a)$ since each update shares of file, the polynomial has change in Equation \eqref{eq-ft}, shares collected across update times cannot be combined to reconstruct $W_n$.

\textbf{Security:} All security keys in the system are one-time keys derived via the DH protocol and the HKDF algorithm. Under the standard computational Diffie-Hellman assumption and the pseudo-randomness of HKDF, each key is computationally indistinguishable from a uniformly random string to any polynomial-time eavesdropper who does not participate in the key agreement. Consequently, the ciphertext reveals no information about the share, and thus no information about any file in the library.

Thus, we have completed the proof of Theorem 1 by verifying all required constraints. %

\section{Example}
\label{sec-example}
The following example is provided as a clear illustration of the $(U, M, N)$ D2D proactive coded caching system.
\begin{example}\rm
Consider a system with a server has $N=4$ files, denoted as $\mathcal{W}=\{W_0,W_1,W_2,W_3\}$. Each file is of size $B$ bits, i.e., $W_{n\in[N]}\in\mathbb{F}_2^B$. It can be divided into equal packets. The server is connected to $U=4$ users, represented by the set $\mathcal{U}=\{0,1,2,3\}$. In this example, we set the scheme parameter $l=1$. Consequently, each user's cache $Z_{u\in \mathcal{U}}$ can store $M=\frac{Nl}{U-l}+\frac{1}{l}=\frac{7}{3}$ files.
\subsection{Placement Phase}
\subsubsection{Prefetching Stage} Each file $W_n\in \mathcal{W}$ is divided into $3$ packets, with each of size $L=\frac{B}{3}$ bits, i.e., \[W_n=\{W_{n,0},W_{n,1},W_{n,2}\}.\] Each file is encoded using the $(1,4,4)$-ramp secret sharing scheme by server. For each file $W_n$, the server generates a random polynomial of degree $3$. For instance, for file $W_0$, the random polynomial is \[f_0^{(0)}(x)=a_{0,0}+W_{0,0}x+W_{0,1}x^2+W_{0,2}x^3\in\mathbb{F}_{2^L}[x],\] where $a_{0,0}\in\mathbb{F}_{2^L}$ is chosen uniformly at random.
For each file $W_n$, the server generates $4$ shares, denoted by $f_n^{(0)}(c_{\mathcal{T}})=S_{n,\mathcal{T}}^{(0)}\in \mathbb{F}_{2^{L}}$, where $c_\mathcal{T}\in \mathbb{F}_{2^{L}}$ is known to all users, $\mathcal{T}\subset [U]$, and $|\mathcal{T}|=1$. Each share has size $F_s=L$ bits. In this example, the cached contents ${Z}_{0,0}^{(0)}$, ${Z}_{0,1}^{(0)}$, ${Z}_{0,2}^{(0)}$, and ${Z}_{0,3}^{(0)}$ are as follows:
\begin{align}\label{eq-example-share}
\begin{aligned}
{Z}_{0,0}^{(0)}=\{S_{0,0}^{(0)},S_{1,0}^{(0)},S_{2,0}^{(0)},S_{3,0}^{(0)}\},\\
{Z}_{0,1}^{(0)}=\{S_{0,1}^{(0)},S_{1,1}^{(0)},S_{2,1}^{(0)},S_{3,1}^{(0)}\},\\
{Z}_{0,2}^{(0)}=\{S_{0,2}^{(0)},S_{1,2}^{(0)},S_{2,2}^{(0)},S_{3,2}^{(0)}\},\\
{Z}_{0,3}^{(0)}=\{S_{0,3}^{(0)},S_{1,3}^{(0)},S_{2,0}^{(0)},S_{3,3}^{(0)}\}.
\end{aligned}
\end{align}

By Algorithm \ref{alg-upkeyd2d-ppk}, every subset of users of size $2$ together generates a master secret key (MSK). Then, the keys of each user are stored in ${Z}_{1,u}$ as follows:
\begin{align}
\begin{aligned}
\label{eq-example-key}
{Z}_{1,0}=\{K_{01},K_{02}, K_{03}\},~{Z}_{1,1}=\{K_{01},K_{12}, K_{13}\},\\
{Z}_{1,2}=\{K_{02},K_{12}, K_{23}\},~{Z}_{1,3}=\{K_{03},K_{13}, K_{23}\}.
\end{aligned}
\end{align}Hence, the cached contents for each user are given by
\begin{align*}
Z_{0}^{(0)}={Z}_{0,0}^{(0)}\cup{Z}_{1,0},~Z_{1}^{(0)}={Z}_{0,1}^{(0)}\cup{Z}_{1,1},\\
Z_{2}^{(0)}={Z}_{0,2}^{(0)}\cup{Z}_{1,2},~Z_{3}^{(0)}={Z}_{0,3}^{(0)}\cup{Z}_{1,3}.
\end{align*}The results show that each user stores $(4+3)\times\frac{B}{3}=MB$ bits, i.e., $M$ files.

\subsubsection{Renewal Stage} For example, consider updating file $W_0$. When performing the $t$-th update, each user $u\in\mathcal{U}$ selects a random constant function \[h_{0,u}^{(t)}(x)=b^u_{0,0} \in\mathbb{F}_{2^L}[x].\] Then, for any singleton set $\mathcal{T}\subset [U]$ with $|\mathcal{T}|=1$, we have $h_{0,u}^{(t)}(c_{\mathcal{T}})=b^u_{0,0}$. The broadcast signals from all users are determined by Equation \eqref{eq-XY}. As illustrated in Table \ref{tab-example}, each row of the table corresponds to the signal broadcast by a user, and each column is representative of the signals that a user is capable of decoding.

\begin{table}[htbp]
\centering
\caption{The signals broadcast and decodable by all users}
\label{tab-example}
\setlength{\tabcolsep}{2.5pt}
\begin{tabular}{cccccc}
\toprule
&&\multicolumn{4}{c}{Decodable signals}\\
\cmidrule{3-6}
&User& $0$&$1$&$2$&$3$\\
\midrule
\multirow{4}{*}{\adjustbox{rotate=270,raise=1.2em}{Broadcast Signals}}&$0$&&$h_{0,0}^{(t)}(c_1)\oplus K_{01}^0$&$h_{0,0}^{(t)}(c_2)\oplus K_{02}^0$&$h_{0,0}^{(t)}(c_3)\oplus K_{03}^0$\\[0.21cm]
&$1$&$h_{0,1}^{(t)}(c_0)\oplus K_{01}^1$&&$h_{0,1}^{(t)}(c_2)\oplus K_{12}^1$&$h_{0,1}^{(t)}(c_3)\oplus K_{13}^1$\\[0.21cm]
&$2$&$h_{0,2}^{(t)}(c_0)\oplus K_{02}^1$&$h_{0,2}^{(t)}(c_1)\oplus K_{12}^0$&&$h_{0,2}^{(t)}(c_3)\oplus K_{23}^0$\\[0.21cm]
&$3$&$h_{0,3}^{(t)}(c_0)\oplus K_{03}^1$&$h_{0,3}^{(t)}(c_1)\oplus K_{13}^0$&$h_{0,3}^{(t)}(c_2)\oplus K_{23}^1$&\\[0.21cm]
\bottomrule
\end{tabular}
\end{table}

Subsequently, each user $u\in\mathcal{U}$ updates the shares of file $W_0$ using Equation \eqref{eq-update}. For instance, user $0$ updates the share $S_{0,0}^{(t-1)}$ to
\begin{align*}
\begin{aligned}
S_{0,0}^{(t)}&=S_{0,0}^{(t-1)}+\sum_{u\in[4]}h_{0,u}^{(t)}(c_{0})\\
&=S_{0,0}^{(t-1)}+\sum_{u\in[4]}b^{u}_{0,0}\\
&=f_0^{(t-1)}(c_{0})+h_0^{(t)}(c_{0})
=f_0^{(t)}(c_{0}) \in\mathbb{F}_{2^L}.
\end{aligned}
\end{align*}Therefore, the updated shares of file $W_0$ are:
\begin{align*}
S_{0,0}^{(t)}=S_{0,0}^{(t-1)}+h_0^{(t)}(c_{0}),~S_{0,1}^{(t)}=S_{0,1}^{(t-1)}+h_0^{(t)}(c_{1}),\\
S_{0,2}^{(t)}=S_{0,2}^{(t-1)}+h_0^{(t)}(c_{2}),~S_{0,3}^{(t)}=S_{0,3}^{(t-1)}+h_0^{(t)}(c_{3}).
\end{align*}

\subsection{Delivery Phase}
Suppose each user has a distinct request, represented by the demand vector $\mathbf{d}=(0,1,2,3)$. Here, $\{K_{\mathcal{L}}^{j}\}_{j\in\{2,3\}}$ are derived from $K_{\mathcal{L}}$ and are available to each user $u\in\mathcal{L}$. The users then transmit the following signals:
\begin{align}
\label{eq-example-delivery}
\begin{aligned}
&X_{0,\mathbf{d}}=\left\{
S_{1,0}^{(t-1)}\oplus K_{01}^{2}, S_{2,0}^{(t-1)}\oplus K_{02}^{2}, S_{3,0}^{(t-1)}\oplus K_{03}^{2},
\right\},\\
&X_{1,\mathbf{d}}=\left\{
S_{0,1}^{(t)}\oplus K_{01}^{3}, S_{2,1}^{(t-1)}\oplus K_{12}^{3}, S_{3,1}^{(t-1)}\oplus K_{13}^{3},
\right\},\\
&X_{2,\mathbf{d}}=\left\{
S_{0,2}^{(t)}\oplus K_{02}^{3}, S_{1,2}^{(t-1)}\oplus K_{12}^{2}, S_{3,2}^{(t-1)}\oplus K_{23}^{2},
\right\},\\
&X_{3,\mathbf{d}}=\left\{
S_{0,3}^{(t)}\oplus K_{03}^{3}, S_{1,3}^{(t-1)}\oplus K_{13}^{2}, S_{2,3}^{(t-1)}\oplus K_{23}^{3},
\right\}.\end{aligned}
\end{align}

The users decode their demanded file as follows. Consider user $0$. User $0$ has $K_{01}$, $K_{02}$, $K_{03}$ in its cache from Equation \eqref{eq-example-key}. Then, it can generate keys $K_{01}^{3}$,$K_{02}^{3}$ and $K_{03}^{3}$. Therefore, it can decode $S_{0,1}^{(t)}$, $S_{0,2}^{(t)}$ and $S_{0,3}^{(t)}$ from $X_{1,\mathbf{d}}$, $X_{2,\mathbf{d}}$, and $X_{3,\mathbf{d}}$, respectively. Moreover, user $0$ already has the share $S_{0,0}^{(t)}$ cached from Equation \eqref{eq-example-share}. Hence, users $0$ can reconstruct the requested file $W_0$. Similarly, user $1$, $2$, and $3$ can reconstruct their respective requested files.

For each user $u\in[4]$, file privacy is ensured by the shares $S^{(t)}_{0,u}$ and $S^{(t-1)}_{0,u}$, which are obtained by evaluating $f^{(t)}_{0}(x)$ and $f^{(t-1)}_{0}(x)$ at $x=c_{u}$, respectively. Here, \[f^{(t)}_{0}(x)=f^{(t-1)}_{0}(x)+h^{(t)}_{0}(x).\] During the renewal stage, users refresh their shares, that is, they update the polynomial. In each update, the polynomial is refreshed, and the reconstruction threshold for file $W_0$ increases by one. This ensures that an attacker from accumulating enough shares to reach the threshold.  Moreover, the security of each transmission is ensured by a unique secure key included in every signal.

Finally, the size of each packet is $\frac{B}{3}$ bits. Each user caches $4+3$ packets and sends $3$ packets. Therefore, the scheme achieves the performance pair $(M,R)=(\frac{7}{3},4)$.
\end{example}

\section{performance analysis}
\label{sec-analysis}
In this section, we present numerical comparisons and, in Subsection \ref{sec-performance}, analyze the performance of the proposed scheme.
\subsection{Numerical results}
\begin{figure}[h]
\centering
\includegraphics[trim=110 270 120 280, clip, width=0.35\textwidth]{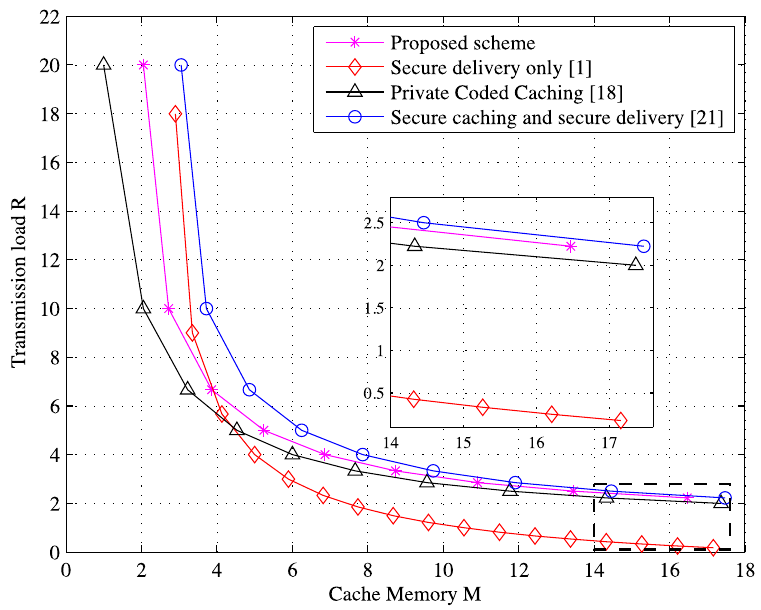}
\caption{Comparison between the required transmission load under different system requirements for $N=U=20$.}\label{fig-example-20}
\end{figure}

In this section, we present numerical results that illustrate the performance of the system. Figure \ref{fig-example-20} shows the numerical evaluation of the proactive caching scheme's performance under different system requirements and $(U,M,N)=(20,M,20)$. Specifically, we compare its performance with that of the system considered in \cite{RPK2017}, which preserves file privacy and uses server-only delivery. As $M$ increases, the difference in transmission load between the two schemes decreases. Additionally, we contrast our scheme with two types of systems: systems with secure delivery only \cite{AS2015} and those that ensure both file privacy and secure delivery \cite{ZIY20192}. The results indicate that the proposed scheme achieves competitive performance. For systems with secure delivery only [1], the smaller size of each file share contributes to a lower overall delivery load. For sufficiently large $M$, the performance loss due to key storage overhead becomes negligible, as expected. Notably, all compared schemes are order-optimal.

\subsection{Performance}
\label{sec-performance}
In this subsection, based on the results in \cite{ZIY20192}, this scheme is shown to be order-optimal when $M\geq 2+\frac{N}{U-1}$; that is, its achievable delivery load $R(M)$ is within a constant multiplicative factor of the information-theoretic lower bound.

Next, we will use the following result, which is Theorem 1 in \cite{ZIY20192}, as key lemma in our proof.
\begin{lemma}[Theorem 1 in \cite{ZIY20192}]\rm
\label{le-ZIY1}
Under centralized placement, for $M'=\frac{Nl}{U-l}+\frac{1}{l}+1$, and $l\in[1,U-1]$, the secure transmission load is upper bounded by
\begin{align*}
\begin{split}
R^*(M')&\leq R^C(M')\\
&\leq \frac{2U(N+M'-1)}{1+(M'-1)U+\sqrt{(1-(M'-1)U)^2-4UN}}\\
&=\frac{U}{l}.
\end{split}
\end{align*}
\end{lemma}

It should be emphasised that the scheme proposed in \cite{ZIY20192} is order-optimal for $M'\geq 2+\frac{N}{U-1}$, i.e.,there exists a constant $c$ such that
\begin{align*}
\frac{R^C(M')}{R^*(M')}\leq c.
\end{align*}

In fact, the value of $R(M')$ decreases as the value of $M'$ increases, i.e., $R^C(M'-1)\leq R^C(M')$. By Lemma \ref{le-ZIY1} and Equation \eqref{eq-Rs}, we can obtain
\begin{align*}
R(M')=R^C(M'-1)\leq R^C(M').
\end{align*}Hence, we have \[\frac{R(M')}{R^*(M')}\leq \frac{R^C(M')}{R^*(M')}\leq c,\] i.e., the proposed scheme is order-optimal.

\section{Conclusion}
\label{sec-coclusion}
We propose a novel D2D proactive coded caching system designed to preserve long-term file privacy. The corresponding caching scheme achieves an efficient memory-load trade-off when the file size is sufficiently large. It has been demonstrated that this scheme is also order-optimal in the large memory.

\bibliographystyle{IEEEtran}
\bibliography{References}

@inproceedings{AS2015,
  author    = {Awan, Z. H. and Sezgin, A.},
  title     = {Fundamental limits of caching in D2D networks with secure delivery},
  booktitle = {2015 IEEE International Conference on Communication Workshop (ICCW)},
  pages     = {464-469},
  year      = {2015}
}

@inproceedings{FHN2014,
  author    = {Flores, W. R. and Holm, H. and Nohlberg, M. and Ekstedt, M.},
  title     = {An empirical investigation of the effect of target-related information in phishing attacks},
  booktitle = {2014 IEEE 18th International Enterprise Distributed Object Computing Conference Workshops and Demonstrations},
  pages     = {357-363},
  year      = {2014}
}

@inproceedings{HJK1995,
  author    = {Herzberg, A. and Jarecki, S. and Krawczyk, H. and Yung, M.},
  title     = {Proactive secret sharing or: How to cope with perpetual leakage},
  booktitle = {Advances in Cryptology-CRYPTO '95, International Cryptology Conference},
  pages     = {339-352},
  year      = {1995}
}

@inproceedings{RPK2016,
  author    = {Ravindrakumar, V. and Panda, P. and Karamchandani, N. and  Prabhakaran, V.},
  title     = {Fundamental limits of secretive coded caching},
  booktitle = {2016 IEEE International Symposium on Information Theory (ISIT)},
  pages     = {425-429},
  year      = {2016}
}

@inproceedings{ZIY20191,
  author    = {Zewail, A. A. and Ibrahim, A. M. and Yener, A.},
  title     = {An optimization framework for secure delivery in heterogeneous coded caching systems},
  booktitle = {2019 53rd Asilomar Conference on Signals, Systems, and Computers},
  pages     = {1232-1236},
  year      = {2019}
}

@inproceedings{BM1985,
  author    = {Blakley, G. R. and Meadows, C.},
  title     = {Security of ramp schemes},
  booktitle = {Workshop on the Theory and Application of Cryptographic Techniques},
  pages     = {242-268},
  year      = {1984}
}

@inproceedings{QCN2024,
  author    = {Qi, C. and Cheng, M. and Niu, X. and Dai, B.},
  title     = {Coded caching with file and demand privacy},
  booktitle = {2024 IEEE International Symposium on Information Theory (ISIT)},
  pages     = {1658-1663},
  year      = {2024}
}

@article{DH1976,
  author  = {Diffie, W. and Hellman, M.},
  title   = {New directions in cryptography},
  journal = {IEEE Transactions on Information Theory},
  volume  = {22},
  number  = {6},
  pages   = {644-654},
  year    = {1976},
  month   = {Nov.}
}

@article{DLW2021,
  author  = {Ding, J. and Lin, C. and Wang, H. and Xing, C.},
  title   = {Communication efficient secret sharing with small share size},
  journal = {IEEE Transactions on Information Theory},
  volume  = {268},
  number  = {1},
  pages   = {659-669},
  year    = {2022},
  month   = {Jan.}
}

@article{ITDW1982,
  author  = {Ingemarsson, I. and Tang, D. and Wong, C.},
  title   = {A conference key distribution system},
  journal = {IEEE Transactions on Information theory},
  volume  = {28},
  number  = {5},
  pages   = {714-720},
  year    = {1982},
  month   = {Sept.}
}

@article{JCM2015,
  author  = {Ji, M. and Caire, G. and Molisch, A. F.},
  title   = {Fundamental limits of caching in wireless D2D networks},
  journal = {IEEE Transactions on Information Theory},
  volume  = {62},
  number  = {2},
  pages   = {849-869},
  year    = {2016},
  month   = {Feb.}
}

@article{M2018,
  author  = {Martinez-Penas, U.},
  title   = {Communication efficient and strongly secure secret sharing schemes based on algebraic geometry codes},
  journal = {IEEE Transactions on Information Theory},
  volume  = {64},
  number  = {6},
  pages   = {4191-4206},
  year    = {2018},
  month   = {Jun.}
}

@article{MN2014,
  author  = {Maddah-Ali, M. A. and Niesen, U.},
  title   = {Fundamental limits of caching},
  journal = {IEEE Transactions on Information Theory},
  volume  = {60},
  number  = {5},
  pages   = {2856-2867},
  year    = {2014},
  month   = {May.}
}

@article{NZV2020,
  author  = {Ni, J. and Zhang, K. and Vasilakos,A. V.},
  title   = {Security and privacy for mobile edge caching: Challenges and solutions},
  journal = {IEEE Wireless Communications},
  volume  = {28},
  number  = {3},
  pages   = {77-83},
  year    = {2020},
  month   = {Dec.}
}

@article{RPK2017,
  author  = {Ravindrakumar, V. and Panda, P. and Karamchandani, N. and Prabhakaran, VM.},
  title   = {Private coded caching},
  journal = {IEEE Transactions on Information Forensics and Security},
  volume  = {13},
  number  = {3},
  pages   = {685-694},
  year    = {2017},
  month   = {Oct.}
}

@article{STC2014,
  author  = {Sengupta, A. and Tandon, R. and Clancy, T. C.},
  title   = {Fundamental limits of caching with secure delivery},
  journal = {IEEE Transactions on Information Forensics and Security},
  volume  = {10},
  number  = {2},
  pages   = {355-370},
  year    = {2015},
  month   = {Feb.}
}

@article{ZIY20192,
  author  = {Zewail, A. A. and Ibrahim, A. M. and Yener, A.},
  title   = {Device-to-device secure coded caching},
  journal = {IEEE Transactions on Information Forensics and Security},
  volume  = {15},
  number  = {},
  pages   = {1513-1524},
  year    = {2019},
  month   = {Sept.}
}

@misc{DW2012,
  author       = {Danaher, B. and Waldfogel, J.},
  title        = {{Reel piracy: The effect of online film piracy on international box office sales}},
  howpublished = {Available at SSRN 1986299},
  year         = {2012}
}

@misc{EMR2025,
  author       = {{Ericsson}},
  title        = {{Ericsson Mobility Report}},
  howpublished = {Ericsson, Stockholm, Sweden, Jun. 2025. [Online]. Available on: \url{https://www.ericsson.com/en/reports-and-papers/mobilityreport/reports/june-2025}}
}

@misc{KE2010,
  author       = {Krawczyk, H. and Eronen, P.},
  title        = {{HMAC-based Extract-and-Expand Key Derivation Function (HKDF)}},
  howpublished = {RFC 5869, Internet Engineering Task Force},
  year         = {2010}
}

@misc{M2024,
  author       = {{Microsoft Corporation}},
  title        = {Microsoft Digital Defense Report 2024},
  howpublished  = {Microsoft Research (2024)},
  year         = {2024},
  month        = {Oct.},
  note         = {[Online]. Available on: \\ \url{https://www.microsoft.com/en-us/security/security-insider/threat-landscape/microsoft-digital-defense-report-2024}}
}

\end{document}